\documentclass[12pt]{article}
\usepackage{graphicx}
\begin{document}

\centerline{\bf \large Sociophysics Simulations IV: Hierarchies of Bonabeau et al}

\bigskip
\centerline{Dietrich Stauffer}

\bigskip
\centerline{Institute for Theoretical Physics, Cologne University, D-50923 K\"oln, Euroland}

\begin{abstract}
The model of Bonabeau et al explains social hierarchies as random: People 
keep a memory of recent fights, and winners have a higher probability to win
again. The question of phase transition and the generalization from square
lattices to networks is reviewed here.
\end{abstract}

\section{Introduction}

Why do some speakers at this 8th Granada Seminar give five lectures and expect 
a million
Euro honorarium, while others are allowed only to present a poster and have to
pay a registration fee. The elites of all times and regions always had 
excellent reasons why they should be on top: They were kings by the grace of 
god, or university professors by their excellent work. Indeed, at the age 
when Albert Einstein wrote about relativity, diffusivity and quantum photo
effect as a technical expert at the patent office in Bern, this author was 
already paid as university assistant, a position never reached by Einstein. 
Thats why this is this author's fourth contribution to these proceedings.

However, bad people \cite{bonabeau} find also other reasons for my position.
The Maxwell-Boltzmann statistics for classical ideal gases gives air molecules
a probability exp($-v^2/2mk_BT$) to have a velocity vector {\bf v}. If one
air molecule has a velocity ten times higher than average, then statistical
physicists see this as a normal though rare random event and do not associate 
any wonderful properties with this molecule. And this is how \cite{bonabeau}
treats us: We got our position in society by accident. These authors are evil
because Bonabeau left academia and Deneubourg gave a talk without a tie at a 
conference where all male speakers were asked months in advance to wear a tie.
Nevertheless we now look at their model, following Rasputin's advice that
one can reject sin only after having studied it. 

\section{Standard Model}

People diffuse on a square lattice filled with density $p$. Whenever a person 
wants to move onto a site already occupied by someone else, a fight erupts 
which is won by the invader with probability $q$ and lost with probability 
$1-q$. If the invader wins, the winner
moves into the contested site whereas the loser moves into the site left free
by the winner; otherwise nobody moves. Each visitor
adds $+1$ to a history parameter $h$, and each loss adds $-1$ to $h$. At each
iteration, the current $h$ is diminished by ten percent, so that roughly only  
the last ten time steps are kept in memory $h$. The probability $q$ for fighter
$i$ to win against fighter $k$ is a Fermi function:
$$ q = 1/[1 + \exp((h_k - h_i)\eta)] \eqno (1) $$
where the free parameter $\eta$ could be unity. Initially everybody starts with
$h = 0$; then $q=1/2$ for all fights. After some time, history $h$ accumulates 
in memory, $q$ differs from 1/2, and the standard deviation $\sigma(t)$ with
$$ \sigma^2 = <q^2> - <q>^2 \eqno (2)$$
measures the amount of inequalities in society at that time step $t$ and is
obtained by averaging over all fights occuring during this iteration $t$. 

This defines the model and the main quantity $\sigma$ to look at. If instead 
one looks at the history of one person and integrates its $h$ over time, after 
sufficiently long times it averages to zero: Who is on top at some time may be
away from the top another time. Real Madrid has shown this to the football 
world: There is always one team winning the Champions League, but it is not
necessarily the same team each year. We find similar examples in the political
powers dominating Europe during the last two-thousand years. 
 
Bonabeau et al \cite{bonabeau} found a phase transition in that for high 
densities the inequalities are strong, and for densities below some threshold
they no longer exist. This effects corresponds to widespread feelings (see
the movie Dances with Wolves) that social hierarchies developed only with
agriculture and cities (but what about the Mongolian empire ?) Unfortunately
that effect was based on an assumption which prevented equilibrium and let the 
$|h|$ go to infinity. When corrected, the phase transition vanished 
\cite{sousa1}. The transition was restored \cite{stauffer} by a feedback loop:
the quantity $\eta$ in eq.(1) was replaced by the $\sigma$ as calculated from
eq.(2) at the previous time step. (For the first 10 time steps, $\sigma$ was
replaced by one.) Then a (first-order) transition was found again, Fig.1.

\begin{figure}[hbt]
\begin{center}
\includegraphics[angle=-90,scale=0.5]{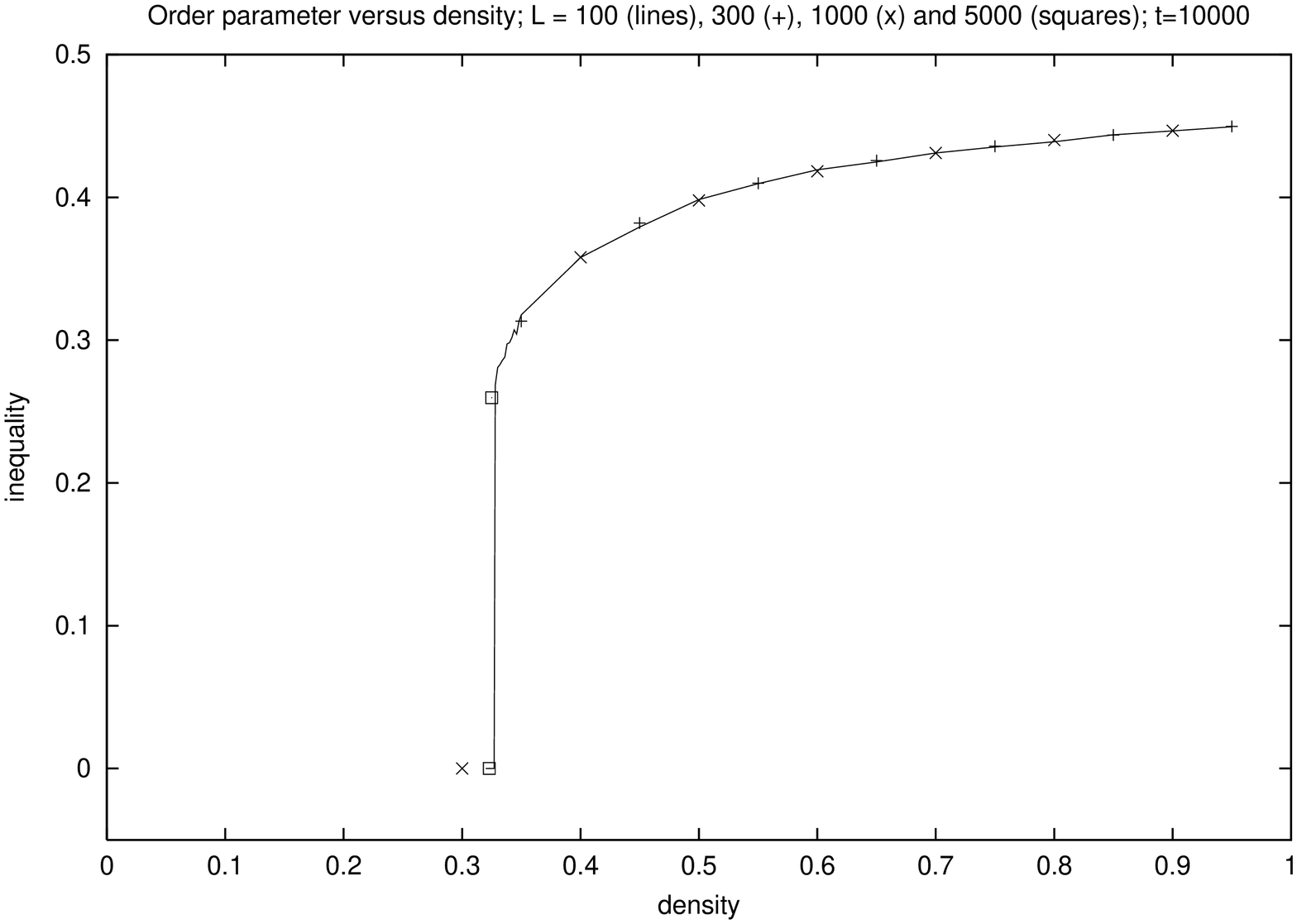}
\end{center}
\caption{ 
}
\end{figure}

\section{Program}

Now follows the Fortran program for this latest version \cite{stauffer}.

\begin{verbatim}
      parameter(p=0.3  ,L= 1000,Lsq=L*L)
      dimension hist(Lsq),latt(Lsq),ipos(Lsq),neighb(0:3)
      real*8 q, qsum, qsu2, factor
      integer*8 ibm,large
      data eta,forget,ibm,maxstep/1.0,.10,1,10000/
     1     large/'7FFFFFFFFFFFFFFF'X/
      print *, p, L, eta, forget, ibm, maxstep
      n=p*Lsq
      fact=Lsq*1.0d0/large
      factor=0.5d0/large
      neighb(0)= 1
      neighb(1)=-1
      neighb(2)= L
      neighb(3)=-L
      ibm=2*ibm-1
      do 1 i=1,n
 1      hist(i)=0
      do 2 j=1,Lsq
 2      latt(j)=0
      do 3 i=1,n
 4      ibm=ibm*16807
        if(ibm.lt.0) ibm=(ibm+large)+1
        j=1+fact*ibm
c       initially random, no two people on one site
        if(latt(j).ne.0) goto 4
        latt(j)=i
 3      ipos(i)=j
c     initialization finished; no dynamics starts
      do 5 itime=1,maxstep
      icount=0
      qsum=0.0d0
      qsu2=0.0d0
        do 6 i=1,n
         hist(i)=hist(i)*(1.0-forget)
          j=ipos(i)
          ibm=ibm*16807
          jnew=j+neighb(ishft(ibm,-62))
          if(jnew.gt.Lsq) jnew=jnew-Lsq
          if(jnew.le.0) jnew=jnew+Lsq
          if(latt(jnew).eq.0) then
c         either new site is empty: move there; or it is occupied: fight
            latt(jnew)=i
            latt(j)=0
            ipos(i)=jnew
          else
            k=latt(jnew)
            qq=eta*(hist(k)-hist(i))
            if(itime.gt.10) qq=qq*sigma
            if(abs(qq).lt.10) then
              q=1./(1.0+exp(qq))
            else
              if(qq.lt.0) q=0.9999
              if(qq.gt.0) q=0.0001
            end if
            icount=icount+1
            qsum=qsum+q
            qsu2=qsu2+q*q
            ibm=ibm*65539
            if(0.5+ibm*factor .lt. q) then
c             now i has won over k and moves
              latt(jnew)=i
              latt(j)=k
              ipos(i)=jnew
              ipos(k)=j
              hist(i)=hist(i)+1.0
              hist(k)=hist(k)-1.0
            else
              hist(i)=hist(i)-1.0
              hist(k)=hist(k)+1.0
            endif
          endif
 6      continue
      qsum=qsum/icount
      qsu2=qsu2/icount
      sigma=sqrt(qsu2-qsum*qsum)
      if(sigma.lt.0.000001) goto 7
 5    print *, itime, sigma, icount
 7    continue
      stop
      end
\end{verbatim}

This program unfortunately violates the Gerling criterion that nobody should
publish more program lines than (s)he has years in life. Thus I start with the
core, after the comment line 41: If the site {\tt jnew} to which agent {\tt i}
wants to move is empty, {\tt latt(jnew)} = 0, then the move is made: the 
position of the agent is now {\tt jnew}, and the occupation variables of the
sites {\tt j, jnew} are interchanged.

Otherwise a fight starts between agent {\tt i} and the present inhabitant
{\tt k}. The probability $q$ from eq.(1) is calculated (with an escape if
the argument of the exponential function is too large) and taken into account
in the averages for $\sigma$, eq.(2). A random integer {\tt ibm}, obtained by 
multiplication with 65539 (or better 16807 as earlier), is compared after
normalization with the probability $q$ of the invader {\tt i} to win; if {\tt i}
wins, again the occupation variables are interchanged, and so are the 
position variables {\tt ipos}; moreover, the history variables {\tt h} are
changes by $\pm 1$. If the invader loses, nobody moves, and only the history
variables are changed in the opposite sense. Then the loop over all {\tt n}
agents ends, $\sigma$ is evaluated and printed out. If $\sigma < 10^{-6}$ or if
the maximum number {\tt maxstep} of iterations is reached, the simulation ends.

\section{Modifications}

If this model \cite{stauffer} leads to hierarchies, then they are symmetric:
There as many people on top as they are on bottom. Reality is different: There 
are few leaders only. This asymmetry was partially reproduced by reducing 
the history counter $h$ by $F$ points, with for example $F=2$, in the case of
a loss, while a victory still increases $h$ by only one point \cite{martins}.

S\'a Martins in that paper \cite{martins} also looked at scale-free networks 
of Barab\'asi-Albert type \cite{ba}. This aspect was studied more thoroughly
by Gallos \cite{gallos} and Sousa \cite{sousa2}. It means we no longer fight 
about territory against whoever sits on the lattice site to which we want to 
move. Instead we fight for power with our acquaintances. And the social network
of acquaintances may be described by scale-free networks, where the number
$k$ of neighbours for each site follows a probability distribution $\propto
1/k^3$ instead of having $k=4$ on the square lattice. Details of the simulations
differ, and so do their results \cite{gallos,martins}, but the sharp phase
transition was recovered. Gallos finds it at a very low concentration $< 0.1$, 
which moreover may decrease towards zero for increasing network size.  
 
A simpler network allows everybody to contact everybody, and also here abrupt 
changes in the amount $\sigma$ of hierarchies were seen \cite{malarz}. Other
cases studied were Erd\"os-R\'enyi random graphs, Watts-Strogatz small-world
networks, and triads where friends of my friends are likely also my own friends
\cite{sousa2}.

\section{Summary}
Even though the model was already published in 1995, it seems to become 
fashionable only now with three independent papers in the first few months 
of 2005 \cite{malarz,gallos,sousa2}. Some crayfish \cite{cray} followed Bonabeau
et al earlier as we physicists. 
 
We thank A,.O. Sousa for a critical reading of the manuscript.

\end{document}